\documentclass[sigplan, screen, authorversion, 10pt]{acmart}
\acmYear{2026}\copyrightyear{2026}
\setcopyright{cc}
\setcctype[4.0]{by}
\acmConference[EUROSYS '26]{European Conference on Computer Systems}{April 27--30, 2026}{Edinburgh, Scotland, UK}
\acmBooktitle{European Conference on Computer Systems (EUROSYS '26), April 27--30, 2026, Edinburgh, Scotland, UK}
\acmDOI{10.1145/3767295.3769373}
\acmISBN{979-8-4007-2212-7/26/04}

\usepackage[normalem]{ulem}

\usepackage{mdframed}
\usepackage{tikz}
\usetikzlibrary{decorations.pathmorphing}
\usepackage{amsmath}
\usepackage{amsfonts}
\usepackage{hyperref}
\usepackage{cleveref}
\crefname{algocf}{alg.}{algs.}
\Crefname{algocf}{Algorithm}{Algorithms}
\usepackage{xspace}
\usepackage{booktabs}
\usepackage{multirow}
\usepackage{stmaryrd}
\usepackage{wrapfig}
\usepackage{enumitem}
\usepackage[frozencache,cachedir=minted-cache]{minted} 
\usepackage{siunitx}

\usepackage[linesnumbered,ruled,noend]{algorithm2e}
\usepackage{subcaption}
\usepackage{graphicx} 
\usepackage{stfloats}
\usepackage{xurl}
\usepackage{circledsteps}
\usepackage{colortbl}
\usepackage{pifont}
\usepackage{balance}
\usepackage{microtype}

\begin{document}

\newcommand{\Comment}[1]{}

\newcommand{\rev}[1]{{\color{red}#1}}
\newcommand{\bhl}[1]{{\color{red}#1}}
\newcommand{\dep}[1]{{\color{red}#1}}

\newcommand{\todoc}[2]{{\textcolor{#1}{\textbf{#2}}}}
\definecolor{darkblue}{rgb}{0.0, 0.0, 0.55}

\newcommand{\todored}[1]{{\todoc{red}{\textbf{[[#1]]}}}}
\newcommand{\todogreen}[1]{\todoc{green}{\textbf{[[#1]]}}}
\newcommand{\todoblue}[1]{\todoc{blue}{\textbf{[[#1]]}}}
\newcommand{\todoorange}[1]{\todoc{orange}{\textbf{[[#1]]}}}
\newcommand{\todobrown}[1]{\todoc{brown}{\textbf{[[#1]]}}}
\newcommand{\todogray}[1]{\todoc{gray}{\textbf{[[#1]]}}}
\newcommand{\todopink}[1]{\todoc{purple}{\textbf{[[#1]]}}}
\newcommand{\todoteal}[1]{\todoc{teal}{\textbf{[[#1]]}}}
\newcommand{\tododarkblue}[1]{\todoc{darkblue}{\textbf{[[#1]]}}}

\newcommand{\todo}[1]{\todored{TODO: #1}}
\newcommand{\osditodo}[1]{\todoorange{TODO: #1}}

\newcommand{\para}[1]{\smallskip\noindent {\bf #1} }

\renewcommand{\todoc}[2]{\relax} 

\newcommand{\finding}[1]{
\begin{mdframed}[linecolor=gray,roundcorner=12pt,backgroundcolor=gray!15,
linewidth=3pt,innerleftmargin=2pt, leftmargin=0cm,rightmargin=0cm,
topline=false,bottomline=false,rightline = false]
#1
\end{mdframed}
}

\newcommand{\cellminipage}[2]{\begin{minipage}{#1}#2\end{minipage}}

\newcommand{\lin}[1]{\todoblue{Lin: #1}} 
\newcommand{\yongle}[1]{\todoteal{Yongle: #1}}
\newcommand{\shangshu}[1]{\todoorange{Shangshu: #1}}

\newcommand{\tool}{C\-\textsc{Sna\-ke}\xspace}
\newcommand{\VC}{Self-Sus\-taining Cascading Failure\xspace}
\newcommand{\VCs}{Self-Sus\-taining Cascading Failures\xspace}
\newcommand{\vc}{self-sus\-taining cascading failure\xspace}
\newcommand{\vcs}{self-sus\-taining cascading failures\xspace}
\newcommand{\Vcs}{Self-sus\-taining cascading failures\xspace}
\newcommand{\Vc}{Self-sus\-taining cascading failure\xspace}

\newcommand{\NumOfSysTested}{five\xspace}
\newcommand{\NumOfNewVC}{15\xspace}
\newcommand{\NumOfNewVCAllowDupe}{17\xspace}
\newcommand{\NumOfNewVCOneDelayAllowDupe}{16\xspace}
\newcommand{\NumOfConfirmedVC}{five\xspace}
\newcommand{\NUMOfConfirmedVC}{Five\xspace}
\newcommand{\NumOfFixedVC}{two\xspace}

\newcommand{\NumOfNewVCNaive}{11\xspace}
\newcommand{\NoNaivePctg}{73\%\xspace}
\newcommand{\NaivePctg}{27\%\xspace}  


\providecommand{\circledtext}[1]{\CircledText[inner xsep=2pt,inner ysep=2pt]{#1}}

\newcommand{\Yes}{\ding{51}}

\newcommand{\interfere}[2]{\texttt{#1}$\rightarrow$\texttt{#2}}

\newcommand{\customrightsquigarrow}[1]{
\begin{tikzpicture}
\tikz \draw [->,
line join=round,
decorate, decoration={
    zigzag,
    segment length=4,
    amplitude=.9,post=lineto,
    post length=2pt
}]  (0,0) -- (#1,0);
\end{tikzpicture}
}
\newcommand{\longrightsquigarrow}{\customrightsquigarrow{0.7}}
\newcommand{\longlongrightsquigarrow}{\customrightsquigarrow{1.0}}

\newcommand{\customrightarrow}[1]{
\begin{tikzpicture}
\tikz \draw [->,
line join=round,
decorate, decoration={
    segment length=4,
    amplitude=.9,post=lineto,
    post length=2pt
}]  (0,0) -- (#1,0);
\end{tikzpicture}
}
\newcommand{\mylongrightarrow}{\customrightarrow{0.7}}
\newcommand{\longlongrightarrow}{\customrightarrow{1.0}}

\makeatletter
\newcommand{\oset}[3][0ex]{%
  \mathrel{\mathop{#3}\limits^{
    \vbox to#1{\kern-2\ex@
    \hbox{$\scriptstyle#2$}\vss}}}}
\makeatother

\newcommand{\OverTextArrow}[1]{\:{\oset[0.8ex]{#1}{\mylongrightarrow}}\,}
\newcommand{\longOverTextArrow}[1]{\:{\oset[0.8ex]{#1}{\longlongrightarrow}}\,}

\newcommand{\EEdgeText}[1]{\texttt{E(#1)}}
\newcommand{\SPlusEdgeText}[1]{\texttt{S}^\texttt{+}\!\texttt{(#1)}}
\newcommand{\EArrow}[1]{\OverTextArrow{\EEdgeText{#1}}}
\newcommand{\SPlusArrow}[1]{\OverTextArrow{\SPlusEdgeText{#1}}}
\newcommand{\longEArrow}[1]{\longOverTextArrow{\EEdgeText{#1}}}
\newcommand{\longSPlusArrow}[1]{\longOverTextArrow{\SPlusEdgeText{#1}}}
\newcommand{\longAnyInterferenceArrow}[1]{\longOverTextArrow{\texttt{E/}\SPlusEdgeText{#1}}}

\newcommand{\DelayExecEdge}[2]{#1$\EArrow{D}$#2}
\newcommand{\DelayIterEdge}[2]{#1$\SPlusArrow{D}$#2}
\newcommand{\FaultExecEdge}[2]{#1$\EArrow{I}$#2}
\newcommand{\FaultIterEdge}[2]{#1$\SPlusArrow{I}$#2}
\newcommand{\CFGEdge}[2]{#1$\OverTextArrow{\texttt{CFG}}$#2}
\newcommand{\CFGBackEdge}[2]{#1$\OverTextArrow{\texttt{ICFG}}$#2}

\Crefformat{section}{§#2#1#3}
\crefformat{section}{§#2#1#3}

\Crefformat{subsection}{§#2#1#3}
\crefformat{subsection}{§#2#1#3}

\Crefformat{subsubsection}{§#2#1#3}
\crefformat{subsubsection}{§#2#1#3}

\title[Detecting Self-Sustaining Cascading Failure via Causal Stitching]{\tool: 
Detecting 
Self-Sustaining 
Cascading Failure 
via Causal Stitching of Fault Propagations
}

\author{Shangshu Qian}
\orcid{0000-0003-2090-7331}
\affiliation{%
  \institution{Purdue University}
  \city{West Lafayette}
  \state{Indiana}
  \country{USA}
}
\email{shangshu@purdue.edu}

\author{Lin Tan}
\orcid{0000-0002-6690-8332}
\affiliation{%
  \institution{Purdue University}
  \city{West Lafayette}
  \state{Indiana}
  \country{USA}
}
\email{lintan@purdue.edu}

\author{Yongle Zhang}
\orcid{0000-0001-5350-5182}
\affiliation{%
  \institution{Purdue University}
  \city{West Lafayette}
  \state{Indiana}
  \country{USA}
}
\email{yonglezh@purdue.edu}

\begin{abstract}

\todo{DO NOT REMOVE: Uncomment the todo relax macro and make sure this line disappears.}

Recent studies have revealed 
that self-sustaining cascading failures 
in distributed systems 
frequently lead to widespread outages, 
which are challenging to contain and recover from. 
Existing failure detection techniques 
struggle to expose such failures prior to deployment, 
as they typically require a complex combination of specific conditions to be triggered. 
This challenge stems from the inherent nature of cascading failures, 
as they typically involve a sequence of fault propagations, 
each activated by distinct 
conditions. 

This paper presents \tool{}, 
a fault injection 
framework 
to expose self-sustaining
cascading failures in distributed systems. 
\tool{} 
uses the novel idea of \textit{causal stitching}, which causally links multiple single-fault injections 
in different test workloads
to simulate complex fault propagation chains.
To identify propagation chains between faults,
\tool{} designs 
a \textit{counterfactual} causality analysis of fault propagations 
-- \textit{fault causality analysis} (FCA): 
FCA compares the execution trace of a fault injection run 
with its corresponding profile run 
(i.e., running the same test without the injection) 
and identifies any additional faults triggered,
which are considered to have a causal relationship with the
injected fault. 

To address the large search space of fault and workload combinations,  
\tool{} employs a \textit{three-phase allocation (3PA)} protocol of test budget 
that prioritizes faults with unique and diverse causal consequences, 
thereby increasing the likelihood of uncovering conditional fault propagations. 
Furthermore, 
to avoid incorrectly connecting fault propagations from workloads with incompatible conditions, 
\tool{} performs a \textit{local compatibility check} 
that \textit{approximately} 
checks the compatibility of 
the path constraints associated 
with connected fault propagations 
with low overhead.

\tool{} has detected \NumOfNewVC bugs that 
resulted in self-sustaining 
cascading failures 
in \NumOfSysTested widely deployed distributed systems, 
\NumOfConfirmedVC of which have been confirmed 
with \NumOfFixedVC fixed. 

\end{abstract}

\maketitle

\section{Introduction}
\label{sec:intro}

Distributed systems are designed to tolerate component failures~\cite{chandra1996unreliable}. 
However, 
recent research has revealed 
\vcs~\cite{li2018pcatch,huang2022metastable,qian2023vicious,guo2013failure}, 
wherein a fault 
propagates across components 
and replicates itself 
through a self-reinforcing loop~\cite{huang2022metastable,qian2023vicious}. 
These failures 
undermine the intended fault tolerance, 
resulting in widespread outages 
with severe consequences. 
For example, 
a \vc happened in Amazon AWS~\cite{AWSOutage2024} 
on July 30th, 2024, 
significantly disrupting core AWS services, 
including AWS Lambda, EC2, and S3. 
The incident also brought down many dependent services, 
such as Whole Foods Supermarket, Amazon Alexa, and Goodreads~\cite{AWSOutage2024Report}.

\begin{figure}[h]
    \centering
    \includegraphics[width=0.7\columnwidth]{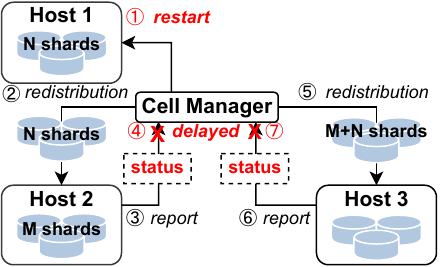}
    \caption{A real-world \vc from Amazon AWS. The Cell (Cluster) Manager manages a cluster of hosts (servers), each storing a set of data shards.}
    \label{fig:aws-exp}
\end{figure}

Such failures often require intricate combinations of conditions 
to manifest~\cite{qian2023vicious}, 
as they typically involve a sequence of fault propagations, 
each activated by distinct conditions. 
For instance, 
as shown in \Cref{fig:aws-exp}, 
the Amazon AWS incident~\cite{AWSOutage2024} 
occurred in a cluster of servers (hosts) 
managed by a cluster (cell) manager, 
with each server hosting a set of data shards. 
During a routine system upgrade, 
the cell manager (\circledtext{1}) restarted some hosts (e.g., Host 1), 
and (\circledtext{2}) redistributed 
their shards to other hosts (e.g., Host 2). 
Due to a latent bug in the load balancer, 
the cell manager mistakenly redistributed 
all \textit{low-throughput shards} 
to a small number of hosts (Host 2). 
Consequently, 
these hosts started 
(\circledtext{3}) sending abnormally large status reports 
that include metadata for all hosted shards. 
The increased size of these reports caused delays 
in both transmission and processing (\circledtext{4}), 
causing the management system to misclassify these servers as unhealthy. 
The management system consequently 
removed these servers from the cluster, 
and (\circledtext{5}) redistributed their shards to other servers, 
which causes the receiving servers (e.g., Host 3) to send large-sized reports (\circledtext{6})
and subsequently be removed from the cluster (\circledtext{7}). 
This failure involves two fault propagations, 
each requiring distinct conditions.
First, for a server removal to propagate and trigger incorrect redistribution 
(\circledtext{1} $\rightarrow$ \circledtext{2} and \circledtext{4} $\rightarrow$ \circledtext{5}), the redistributed shards must have low throughput. 
Second, for incorrect redistribution to induce report delay and subsequent server removal
(\circledtext{2} $\rightarrow$ \circledtext{4} and \circledtext{5} $\rightarrow$ \circledtext{7}), 
the report size (or, the number of affected shards) must be sufficiently large.

Existing fault injection techniques 
\cite{wu2024efficient,li2018pcatch,crashfuzz, joshi2011prefail,natella2016assessing,
marinescu2009lfi,
alquraan2018analysis,
alvaro2015lineage,
chen2020cofi,
gunawi2011fate,
ju2013fault,
lu2019crashtuner,
mohan2018finding,
li2019efficient,
sun2022automatic,
meiklejohn2021service,
zhang20213milebeach,
chen2023push,
jepsenio} 
fall short of 
exposing such \vcs before deployment, 
because they 
only inject faults into 
a limited set of manually selected (e.g., stress tests) 
or synthetically crafted workloads (e.g., benchmark workloads), 
which often \textbf{lack the required combination of conditions}. 
Conversely, 
if a workload satisfying both of the previously described conditions 
had been exercised, 
a traditional fault commonly used in existing fault injection frameworks 
-- a server crash -- 
would have exposed the AWS \vc before deployment. 
Missing triggering conditions is a known challenge to effective fault injection testing~\cite{li2021challenges, chen2023push}. 
Without prior knowledge of the targeted bug, it is exceptionally hard for the developers to manually create test workloads or rules for workload generators~\cite{kiskis1996synthetic, han1995doctor, marinescu2009lfi, joshi2011prefail} that meet all necessary conditions.

To tackle 
this challenge, 
we propose \textbf{Causal Stitching} to simulate complex fault propagation under complex conditions by \textbf{causally linking multiple single-fault injections in different workloads}. 
Each injection uncovers one step in the fault propagation chain. 
Our insight is that
triggering one step in the chain requires less stringent conditions than triggering the whole chain, increasing the likelihood of achieving this through fault injection into existing workloads, such as the integration tests shipped with the system.

To illustrate, consider the two faults involved in the Amazon AWS \vc: 
node loss ($f_1$) and imbalanced shard redistribution ($f_2$). 
Their causal relationships can be identified through separate fault injection runs. 
Specifically, in a test case $t_1$ with low-throughput shards (condition $c_1$), 
injecting $f_1$ results in the triggering of an additional fault $f_2$, establishing the causal relationship $f_1 \rightarrow f_2$. 
Conversely, in a test case $t_2$ with a large number of shards hosted on individual nodes (condition $c_2$) , 
injecting $f_2$ (simulated via induced delay) 
leads to the manifestation of $f_1$, 
indicating a reverse causal relationship  $f_2 \rightarrow f_1$. 
The \vc can be reconstructed by linking the causal relationship among the two faults 
as long as the workload and conditions in tests $t_1$ and $t_2$ are \textit{compatible}, resulting in a causal cycle of $f_1 \rightarrow f_2 \rightarrow f_1$.

We could not apply this idea to the AWS incident due to the lack of access to the source code and test suite. 
However, our tool -- \tool -- demonstrated its capability to detect a similar, previously unknown \vcs in HBase~\cite{apache-hbase}, an open-source distributed database (\Cref{sec:eval:case-study}). 
Notably, HBase's test suite does not contain a single workload that satisfies all the necessary triggering conditions; 
instead, \tool identified and connected causal relationships across multiple test cases to reveal the failure.

To obtain the causal relationships between faults inside the system, we propose a \textit{counterfactual} causality analysis of fault propagations -- \textbf{fault causality analysis (FCA)}.
FCA compares the execution trace of a fault injection run with its corresponding profile run (i.e., running the same test without the injection) and identifies any additional faults triggered, which are considered to have a causal relationship with the injected fault.

A major challenge faced by \textbf{Causal Stitching} 
is the vast number of combinations of faults and workloads. 
To efficiently explore the causal relationship between faults 
in such a massive search space, 
\tool uses a \textbf{three-phase allocation (3PA) protocol} during test execution to maximize the number of causal relationships discovered under a fixed test budget. 
Intuitively, 
3PA prioritizes injecting faults with unique and diverse causal consequences. 
This prioritization is guided by 
an \textit{adaptive weighting algorithm}, 
which leverages runtime feedback from prior fault injection runs 
to estimate the potential of each fault to 
uncover new causal relationships, particularly conditional propagations.
In addition, 
in the subsequent cycle detection phase, 
\tool utilizes a \textbf{parallel beam search} 
to construct propagation chains, 
also applying this prioritization principle 
to favor faults with unique and diverse causal consequences.

Another challenge is the risk of linking causal relationships 
discovered in workloads with incompatible conditions. 
For example, 
suppose the causal relation 
$f_1 \rightarrow f_2$ is observed in test $t_1$ under a condition $c_1$, 
while $f_2 \rightarrow f_1$ in test $t_2$ happens under the negation of $c_1$, 
linking $f_1 \rightarrow f_2$ and $f_2 \rightarrow f_1$ into a single causal cycle is unsound, as the conditions required for each step are mutually exclusive. 
To reduce the risk while minimize the overhead, \tool approximates such symbolic constraints with the fault's local program trace, including branch evaluation results and the call stack.
\tool performs a \textbf{local compatibility check} with such approximated constraints 
before linking causal relationships, 
and assumes that if the traces leading to the same fault are similar in different tests (e.g., traces leading to $f_2$ in $t_1$ and $t_2$), the compatibility between tests exists.

In summary, this paper makes the following contributions:

\begin{itemize}[leftmargin=3ex, noitemsep, topsep=0pt]
\item We propose \textbf{Causal Stitching}, a fault injection technique 
that detects complex fault propagations by causally linking multiple single-fault injections across workloads. 
\item To overcome the challenge of a massive search space, 
we propose a test budget allocation protocol (\textbf{3PA}) and a \textbf{parallel beam search} cycle detection algorithm, both of which prioritize faults with unique and diverse causal consequences. 
\item To reduce the risk of invalid linking of causal relationships, 
we implement a \textbf{local compatibility check} to eliminate incompatible conditions. 
\item We implement the fault injection and analysis framework, \tool. 
We evaluate \tool on the latest versions of \NumOfSysTested popular open-source distributed systems, i.e., HDFS 2.10.2~\cite{apache-hdfs}, HDFS 3.4.1~\cite{apache-hdfs3}, HBase 2.6.0~\cite{apache-hbase}, OZone 1.4.0~\cite{apache-ozone}, and Flink 1.20.0~\cite{apache-flink}, where \tool detects \NumOfNewVC new \vcs, \NumOfConfirmedVC of which have been confirmed with \NumOfFixedVC fixed by developers. 
\tool's source code is available at \url{https://github.com/Purdue-PFL/CSnake}.
\end{itemize}

\section{Problem Statement}
\label{sec:def}

To introduce our problem definition, 
we first present 
a model of fault injection experiments of cascading failures 
extending the model of general fault injection experiments 
\cite{natella2016assessing}, 
as well as our fault model and causality model.

\begin{figure}[h]
    \centering
    \includegraphics[width=0.8\columnwidth]{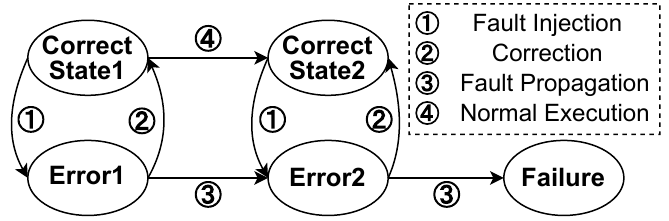}
    \caption{Model of fault injection experiments of cascading failures, 
    characterized by different system states.}
    \label{fig:cascade-fault}
\end{figure}

\textbf{Model of Fault Injection}. As shown in \Cref{fig:cascade-fault}, 
fault injection experiments 
are modeled using transitions between system states. 
During normal system execution (\circledtext{4}), 
a \textit{fault} can be injected (\circledtext{1}) at different time points (Correct State1 and State2) to trigger an \textit{error} (Error1 and Error2). 
The injection can be performed in multiple experiments with different \textit{workloads} (i.e., system inputs). 
An error 
could be masked or corrected (\circledtext{2}) 
by fault tolerance mechanisms such as 
replication~\cite{hdfs_ha} and recomputation~\cite{Zaharia:2012}. 
Under special conditions, 
it could propagate (\circledtext{3}) 
and corrupt other parts of the system state to \textit{cause} additional errors. 
Such propagations (e.g., Error1 $\rightarrow$ Error2 and Error2 $\rightarrow$ Failure) may require distinct \textit{conditions} to be \textit{activated}. A \textit{condition} is a logical predicate over the system state. 

Compared to the general fault injection model~\cite{natella2016assessing}, 
our model captures the causal relationships between faults across \textbf{multiple} fault injection experiments.  
Specifically, in one experiment, a fault ($f_1$) injected into Correct State1 triggers Error1 and subsequently \textit{causes} Error2 (without leading to Failure). 
In a separate experiment, a fault ($f_2$) injected into Correct State2 triggers Error2 and subsequently \textit{causes} Failure. 
The model captures the possible fault propagation Error1 $\rightarrow$ Error2 $\rightarrow$ Failure, provided that the activation conditions for each causal relationship are \textit{compatible}. 
The definitions of \textit{fault}, \textit{causality}, and \textit{compatibility} are provided below. 

\textbf{Fault Model}. 
Traditionally, 
\textit{error} refers to an incorrect system state, 
while \textit{fault} refers to its cause, 
such as software bugs and hardware faults~\cite{natella2016assessing}. 
In the remainder of this paper, 
we use \textit{fault} and \textit{error} interchangeably, 
because we perform a specific type of fault injection 
-- \textit{software-implemented fault injection}~\cite{natella2016assessing}, 
which injects the \textit{effects} of a fault, 
such as exceptions and delays, to simulate \textit{errors} directly and accelerate the experiment, 
instead of injecting the actual faults. 

\textbf{Causality Model}. 
To capture the causal relationship between \textit{faults} (\textit{errors}), 
we use \textit{counterfactual causality}: 
$f_1$ is a counterfactual cause of $f_2$ if and only if $f_2$ would not occur unless $f_1$ occurs. 
According to the ladder of causation~\cite{pearl2018book}, 
\textit{counterfactual causality} offers 
the strongest bond between cause and effect. 
To the best of our knowledge, 
although recently utilized in root cause localization~\cite{fariha2020causality,johnson2020causal}, 
\textit{counterfactual causality} has not been used in fault injection 
to analyze how faults propagate. 
Existing analysis of failure propagation~\cite{li2018pcatch,li2024exchain} 
utilizes the happens-before relationship~\cite{li2018pcatch} 
and program dependencies~\cite{li2024exchain}, 
both of which are weaker forms of causality.  

\textbf{Problem Definition}. 
Given a set of workloads $W = \{w_1, w_2, ...\}$, 
our main goal is to identify the causal relationships between faults 
($f_1 \rightarrow f_2$) across workloads, 
and connect \textit{compatible} 
identified causal relationships 
to detect \textit{cycles}, 
where a fault causes itself through a chain of connected causal relationships.

Because the causal relationships may be identified 
in different workloads (e.g., $f_1 \rightarrow f_2$ identified in $w_1$ and $f_2 \rightarrow f_3$ in $w_2$), 
blindly connecting them could result in invalid causal chains 
when the workloads are \textit{incompatible} -- the \textit{conditions} required to activate $f_1 \rightarrow f_2$ in $w_1$ and $f_2 \rightarrow f_3$ in $w_2$ are mutually exclusive. 
Therefore, 
we require the connected causal relationships to be \textit{compatible}, 
meaning the conjunction of their activating conditions is satisfiable.

For simplicity of the paper, in a causal relationship $f_1 \rightarrow f_2$ identified from $w_1$, we name $f_2$ the \textit{interference} of $f_1$ on the system due to the fault injection. The causal relationship also forms a \textit{fault propagation chain} of length 1 from $f_1$ to $f_2$.

\section{\tool Overview}
\label{sec:overview}

\begin{figure*}[tbh]
    \centering
    \includegraphics[width=0.99\linewidth]{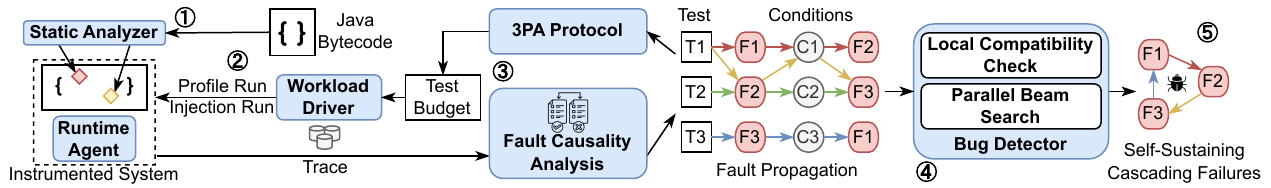}
    \caption{Overview of \tool. Blue boxes are components of \tool.}
    \label{fig:tool-overview}
\end{figure*}

\tool operates by identifying causal relationships between faults in distributed systems and linking them to form potential cycles. It simulates 
chains of fault propagations 
in complex workloads quasi-statically by combining the causal relationships derived from fault injections on simpler workloads.

At the core of \tool is a counterfactual causality analysis. 
Counterfactual causality analysis involves conducting two experiments: 
one with a fault injected to observe its impact on the system execution (referred to as the \textit{interference}), 
and the other without the fault injected to observe the system's execution (referred to as the counterfactual -- i.e., what would the system execution be had the fault not been injected). 
Specifically, we compare the execution trace of a fault injection run with its corresponding profile run 
(i.e., running the same test without the injection) 
and identify any additional faults triggered, 
which are considered to have a counterfactual causal relationship 
with the injected fault.

\textbf{Workflow}. 
\Cref{fig:tool-overview} illustrates the workflow of \tool.
(\circledtext{1}) Given the Java bytecode of the target system, 
\tool's \textit{static analyzer} identifies 
system-specific fault injection points, 
and instruments fault injection and monitoring hooks. 
At runtime, 
\tool's \textit{runtime agent} 
performs fault injection and monitoring 
when a hook is encountered.

During fault injection experiments (\circledtext{2}), 
a \textit{workload driver} selects combinations of faults and workloads 
(integration tests shipped with the target system) 
to execute both the fault injection run 
and the corresponding fault-free profile run (counterfactual experiment).
Note that, 
for each fault, 
\tool only uses the set of tests that 
can reach the fault during execution 
in fault injection runs.

Next, the \textit{fault causality analysis} module (\circledtext{3}) compares the execution traces from the profile and injection runs 
to identify counterfactual causal relationships 
-- i.e., \textit{fault propagations}. 
These causal relationships are used by the 
\textit{three-phase allocation (3PA) protocol} 
to guide future workload and fault selection 
via feedback to the workload driver.

The discovered causal relationships are also forwarded to \tool's \textit{bug detector} (\circledtext{4}), 
which connects \textit{compatible} relationships and 
performs a \textit{beam search} for cycle detection 
to identify \vcs (\circledtext{5}).

In the following sections, 
we first explain 
(\Cref{sec:fia}) 
how to instrument faults 
and perform the core counterfactual causality analysis, 
and then (\Cref{sec:alloc}) how to efficiently allocate testing resources, 
and finally (\Cref{sec:vc-detection}) how to connect compatible causal relationships and detect cycles.

\section{Fault Causality Analysis}
\label{sec:fia}

This section details 
the types of faults to inject (\Cref{sec:fia:faults-to-inject}), 
the fault injection process (\Cref{sec:fia:how-to-inject}), 
and the approach for establishing causal relationships between faults (\Cref{sec:fia:establish-causality}). 

\subsection{What faults to inject?} 
\label{sec:fia:faults-to-inject}

A recent study~\cite{qian2023vicious} 
on \vcs from open-source systems 
reveals that \vcs are often caused by 
functional and performance interferences 
between requests and error handlers. 
Functional interferences typically manifest 
as exceptions and errors captured by system-specific error detectors, 
and performance interferences typically manifest as contention. 
Following this observation, 
\tool injects three types of faults. 

\textbf{Exception}. 
A target system can encounter three types of exceptions 
during its execution: 
(1) system-specific exceptions 
-- exceptions thrown explicitly inside the target system's source code, 
(2) library function exceptions 
-- exceptions thrown by a library or native function, 
(3) unchecked (runtime) exceptions 
-- exceptions that are not required to be explicitly handled 
(by try/catch blocks in JVM-based systems) 
or declared explicitly in function signatures. 

\tool injects system-specific exceptions 
and library function exceptions, 
because they occur at limited, explicitly-declared program locations, 
while unchecked exceptions can happen at arbitrary locations. 
Note that \tool still injects an unchecked exception 
if it is explicitly thrown in the system's code 
-- e.g., an \texttt{IllegalArgumentException} is often thrown 
when a request with invalid input is received. 
In practice, 
\tool filters out reflection-related and
security-related exceptions, 
as they tend to terminate or hang the system 
rather than propagate. 
\tool also ignores exceptions 
only reachable from tests.

\textbf{Contention}. 
Though contention can happen at arbitrary program locations, 
\tool only considers loops for contention injection 
to simulate resource-intensive tasks causing contention, 
due to their association with performance issues discovered by existing studies~\cite{li2018pcatch, jin2012understanding, qian2023vicious}.
\tool uses iteration count of the workload-related loops to capture contention induced by increased workload (\Cref{sec:fia:establish-causality}), similar to the approach adopted by \citet{li2018pcatch}. 

\tool adopts a loop scalability analysis 
to filter out loops unlikely to cause performance issues. 
\tool excludes loops with a constant upper bound on their iteration count, detected by a best-effort data flow analysis to track the loop guard condition to a constant. 
In addition, 
\tool ranks the size of code reachable recursively starting from a loop 
and identifies loops involving I/O operations, 
to exclude loops that have a short execution (i.e., lowest ranked 10\% of loops) and do not perform I/O. 

\textbf{System-Specific Error}. 
Many distributed systems implement system-specific error detectors, 
such as status checks and health monitoring, 
without using exceptions. 
Such system-specific error detectors 
are often implemented as functions with a boolean return value. 
For example, in HDFS, a NameNode thread checks 
if a DataNode's heartbeat is received within a timeout limit 
using a function \texttt{node.isStale()}.

\tool filters out boolean-returning functions 
if they are unlikely to be a system-specific error detector. 
For example, 
\tool filters out such boolean-returning functions in JDK libraries, 
such as \texttt{contains()} function of Java collection type objects 
(detailed filtering criteria in \Cref{sec:impl-detail}). 
\tool's fault filtering criteria 
is designed to be conservative
to avoid meaningful fault being skipped accidentally. 
The injection points that do not impact system execution will be automatically deprioritized during fault injection experiments 
by the test budget allocation protocol (\Cref{sec:alloc}).

\subsection{How to inject faults?}
\label{sec:fia:how-to-inject}

\begin{figure}[tb]
\centering
\footnotesize
\begin{minted}[
  xleftmargin=2ex, xrightmargin=1ex, frame=lines, fontsize=\footnotesize, linenos, 
  autogobble, numbersep=1ex, framesep=1mm, breaklines]
{java}
for (int i = 0; i < 10; i++) { // Loop 1
  sock.read(); // TP1
  if (condition1) {// TP2; MP1
    throw new IOException();
  }
  for (int j = 0; j < 10; j++) { // Loop 2
    if (foo()) { // TP3; MP2
      throw new IOException();
    }
  }
}

boolean foo() { // NP1
  fis = new FileInputStream(path); // TP 4
  if (fis.read() == -1) { // MP 3; TP 5
    return false;
  }
  return true;
}
\end{minted}
\caption{Pseudo-code demonstrating the injection and monitor points. ``TP'' means throw point and ``NP'' negation point. 
``MP'' means monitor point (used and explained in \Cref{sec:chain-forming:approx-check}).}
\label{fig:injection-point-exp}
\end{figure}

\noindent
\tool's static analyzer 
identifies locations to inject faults 
and instruments fault injection hooks statically. 
The hook transfers the control to \tool's runtime agent. 
During fault injection experiments, 
\tool dynamically injects faults 
when corresponding hooks are encountered.  
We use 
a Java-style pseudo-code with eight fault locations
(shown in \Cref{fig:injection-point-exp})
to explain how fault injection is performed.

\textbf{Exception Injection}. 
For system-specific exceptions, 
as they are typically thrown inside an if-statement 
(lines 4 and 8), 
\tool injects 
a one-time throw of the same exception 
when the 
if-statement (lines 3 and 7, referred to as \textit{Throw Point}) is reached. 
For library function exceptions, 
\tool injects a one-time throw of the exception 
declared in the function signature 
at the invocation site of these functions 
(lines 2, 14, and 15). 
Exceptions are constructed at runtime 
using their simplest constructor.

\textbf{Contention Injection}. 
For potentially non-scalable loops 
identified by \tool (e.g., line 1 and 6), 
\tool 
injects a predefined amount of \textbf{spinning delay} 
into each loop iteration 
before the first line of the loop body executes 
to simulate potential contention induced 
by executing a large number of iterations of this loop. 
Due to the way we inject contention, 
we refer to such injection as \textit{delay injection}.

Each delay injection is performed seven times with varying length (100ms to 8s) to maximize the discovery of causal relationships between faults. 
The system is configured with \textit{reduced timeouts} (10-20 seconds) to make it more sensitive to additional workload.

The delay and timeout values are determined empirically to maximize the impact of delay injections. We first lower system timeout configurations to a safe range of 10-20s via trial and error, which ensures normal system functionality being preserved (i.e., all unit and integration tests still pass).
Based on this threshold, we select seven static delay values between 100ms and 8s that are likely to cause timeouts when injected repeatedly inside loops. 
We identify relevant configuration items using the keywords of ``timeout'' and ``interval''. 
The timeout adjustment process typically takes a student 30 minutes per system, and are not essential for \tool's effectiveness, as delay-related causal relationships can still be observed with default settings.

\textbf{System-Specific Error Injection}. 
Because system-specific errors 
are captured by system-specific error detectors, 
whose return values are boolean typed, 
\tool negates the return value of these functions 
(which is referred to as \textit{Negation Point}) 
before the execution returns to its caller 
to simulate the effect of such faults 
(\texttt{foo()} at line 13). 
We refer to the injection of system-specific error as a \textit{negation injection}.

\subsection{How to establish causal relationships?} 
\label{sec:fia:establish-causality}

\label{sec:fia:interference-analysis}

\tool records encountered faults during the injection run 
and profile run, 
and compares them to identify additional faults triggered 
to establish causal relationships. 
We categorize the encountered faults into the following categories: 

\begin{enumerate}[wide, labelwidth=!, labelindent=\parindent, noitemsep, topsep=0pt]
\item \textbf{Execution Trace Interference}. 
For \textbf{exceptions}, 
\tool monitors whether the throw statement is reached. 
For \textbf{system-specific errors}, 
\tool monitors whether 
the return value of the error-detector function is negated. 

\item \textbf{Iteration Count Interference}. 
For \textbf{contention}, 
\tool monitors whether any loop's iteration count \textit{statistically} increases compared to the profile run, 
as an indicator for contention. 
The loop iteration count serves as a good indicator for the amount of workload processed~\cite{li2018pcatch}. 
We use one-sided t-test~\cite{student1908probable} with a $p$-value of 0.1 for statistical significance.

\end{enumerate}

\tool executes each profile run and injection run five times to reduce the impact of \textbf{non-determinism} in the system execution. This also allows us to use statistical tests on loop iteration counts.

Injecting $f_1$ may trigger multiple additional faults, stemming from both interference types. After all the injection runs for $f_1$, \tool collects a list of additional faults triggered $[f_2, f_3, ..., f_n]$.

\textbf{Nested and Consecutive Loops.}
One special case of iteration count interference is the handling of nested and consecutive loops in the system. 
Three types of compositions of workload related loops exist: 1) independent loops, 2) nested loops, and 3) consecutive loops. 
We observe that 2) nested and 3) consecutive workload-related loops are often used in batch processing. 
In such loops, 
a delayed sub-request 
can propagate delays to its parent request (the batch) 
and the next sub-request in the same batch.  
For example, in a batched RPC request, a delayed sub-request can time out the entire call as well as the next sub-request.

\begin{figure}
\centering
\footnotesize
\begin{minted}[
  xleftmargin=2ex, xrightmargin=1ex, frame=lines, fontsize=\footnotesize, linenos, 
  autogobble, numbersep=1ex, framesep=1mm, breaklines]
{java}
while (shouldRun()) { // Loop 1
    resp = sendHeartBeat();
    // Loop 2
    for (DataNodeCommand cmd: resp.getCommands()) {
        processCommand(cmd);
    }
    // ...
    List<BlockReport> reports = new ArrayList<>();
    // Loop 3
    for (Map.Entry kv: perVolumeBlocks.entrySet()) { 
        reports.add(convertFormat(kv));
    }
    // ...
    for (BlockReport report: reports) {
        nnRpcStub.blockReport(report);
        // ...
    }
}
\end{minted}
\caption{Pseudo-code demonstrating delay in nested loops, code simplified from \texttt{BPServiceActor} in HDFS.}
\label{fig:nested-loop-delay}
\end{figure}

\tool identifies the parent-child loop pairs 
to establish this causal relationship and expand the impact scope of contention injection.
\Cref{fig:nested-loop-delay} shows the pseudo-code demonstrating contentions in the nested loops. 
If injection $f_1$ increases iterations in loop 2 ($L_2$), its parent ($L_1$) and sibling ($L_3$) can also be affected.
We use \CFGBackEdge{$L_2$}{$L_1$} to represent the former (``I'' for ``Inverse'') and \CFGEdge{$L_1$}{$L_3$} to represent the latter. 

\textbf{Summary for Fault Causality}. 
\Cref{tab:fault-interference-summary} summarizes all the six causal relationships between different types of faults. 
The first four are differentiated 
by the injected fault and the additional fault triggered. 
The last two are special causal relationships handling nested loops, extending the impact range of contentions.

\begin{enumerate}[leftmargin=3ex, noitemsep, topsep=2pt]
\item \DelayExecEdge{$f_1$}{$f_2$}: Injecting a delay $f_1$ into a loop causes an additional exception or negation $f_2$. Delay $f_1$ has an execution trace interference $f_2$ on the system (hence the name \texttt{E(D)}).

\item \DelayIterEdge{$f_1$}{$f_2$}: Injecting a delay $f_1$ into a loop causes an additional delay ($f_2$) in another loop. Delay $f_1$ has an iteration count interference $f_2$ on the system (name \texttt{S$^{+}$} indicates a statistically significant increase).

\item \FaultExecEdge{$f_1$}{$f_2$}: Injecting an exception or negation $f_1$ causes an additional exception or negation $f_2$. 

\item \FaultIterEdge{$f_1$}{$f_2$}: Injecting an exception or negation $f_1$ causes an additional delay ($f_2$) in another loop.

\item \CFGBackEdge{$f_2$}{$f_2'$}: $f_2$ is an additional delay caused by the injection $f_1$, affecting its parent loop ({$f_2'$}).

\item \CFGEdge{$f_2'$}{$f_2''$}: Following \CFGBackEdge{$f_2$}{$f_2'$}, the parent loop delay ($f_2'$) further affects the sibling loop ($f_2''$).
\end{enumerate}

\begin{table}[t]
\centering
\footnotesize

\caption{Summary of the causal relationships between faults.}

\begin{tabular}{ccc}
\toprule
\textbf{Type} & \textbf{Injected Fault} & \textbf{Additional Fault} \\
\midrule

\DelayExecEdge{}{} & Delay & Exception; Negation \\
\DelayIterEdge{}{} & Delay & Delay \\
\FaultExecEdge{}{} & Exception; Negation & Exception; Negation \\
\FaultIterEdge{}{} & Exception; Negation & Delay \\

\midrule 
\CFGBackEdge{}{} & \multicolumn{2}{c}{$f_1\longSPlusArrow{D/I}$\CFGBackEdge{$f_2$}{$f_2'$} only} \\
\CFGEdge{}{} & \multicolumn{2}{c}{$f_1\longSPlusArrow{D/I}$\CFGBackEdge{$f_2$}{$f_2'$}\CFGEdge{}{$f_2''$} only} \\

\bottomrule

\end{tabular}
\vspace{-2mm}

\label{tab:fault-interference-summary}
\end{table}

\section{Test Budget Allocation}
\label{sec:alloc}

A challenge faced by the approach of \textbf{causal stitching} 
is the vast number of combinations of faults and workloads. 
For example, HDFS has about 3,000 source code locations 
where exceptions can be thrown 
and 2,000 loops where delays can be injected 
and thousands of integration tests. 
Combining injected faults 
with different workloads can result in
tens of millions of possible fault injection scenarios. 
In this section, 
we explain \tool's \textbf{three-phase allocation (3PA) protocol} of test budget, 
which maximizes the discovery of causal relationships 
within a fixed test budget. 
\footnote{A formal definition of the 3PA protocol is in \Cref{app:3pa}.}

\subsection{Principles of Test Allocation}

The 3PA protocol 
leverages two principles  
to prioritize faults based on 
the uniqueness and diversity of their causal impact: 
1) injecting \textit{causally equivalent} faults to diverse workloads, 
and 2) extending \textit{conditional} causal relationships.

\textbf{Causally Equivalent Fault}. 
Since 
different faults (e.g., $f_1$ and $f_2$) 
may trigger similar interferences (i.e., additional faults), 
we consider such faults ($f_1$ and $f_2$) \textit{causally equivalent} with respect to their consequences. 
For example, in a try-catch block, 
multiple exceptions in the try block 
might be handled by the same catch block, 
in which triggers an additional fault. 
In such cases, 
it is redundant to inject \textit{causally equivalent} faults 
into the same workload. 
Instead, 
\tool prioritizes to inject them in different workloads 
in order to maximize the discovery of diverse fault propagations. 

\textbf{Conditional Causal Relationship}. 
Because
\vcs 
are often formed by 
fault propagations activated by specific conditions, 
\tool 
allocates test budget 
to increase the probability 
to extend \textit{conditional} causal relationships between faults. 
\tool detects 
such \textit{conditional} causal relationships 
by checking if a fault causes different faults in different workloads. 
For example, 
if $f_1$ always causes $f_2$ in different workloads, $f_1 \rightarrow f_2$ is considered \textit{unconditional}, 
while if $f_1$ causes $f_2$ in one workload but $f_3$ in another, 
then $f_1 \rightarrow f_2$ and $f_1 \rightarrow f_3$ are considered 
\textit{conditional}. 
\tool prioritizes to allocate test budgets 
to trigger the resulting faults ($f_2$ and $f_3$) 
of such conditional causal relationships.

\subsection{Three-Phase Allocation Protocol}
\label{sec:alloc:3pa}

Following the above principles, 
\tool allocates the test budget in three phases. 
In phase one (\textbf{causally equivalent fault detection}), 
\tool conducts a preliminary exploration 
by running each fault injection against one test workload, 
and clusters faults causing similar interferences 
in order to detect \textit{causally equivalent} faults. 
In phase two (\textbf{causality exploration}), 
\tool picks a fault from each set of \textit{causally equivalent} faults 
and injects it into diverse workloads 
to explore its causal relationships with other faults. 
In phase three (\textbf{conditional-causality-guided causality extension}),  
\tool uses the result from phase two to detect 
\textit{conditional} causal relationships, 
and prioritizes to inject faults that 
can extend such \textit{conditional} causal relationships. 
The total test budget is calculated 
using the number of fault locations in each system, 
currently as $4 \times |\mathbb{F}|$, 
where $\mathbb{F}$ is the set of faults.
Phase one makes up the first 25\% of the budget. Phase two  gets 50\% of the budget, while phase three uses the remaining 25\%.

\textbf{Phase One: Causally Equivalent Fault Detection}. 
In this phase, 
for each fault $f_i$, 
\tool picks the workload $t_{i_1}$ that reaches $f_i$'s program location 
and has the highest code coverage. 
The rationale is that 
such a test is more likely to 
reveal the most diverse interferences of $f_i$. 
Different faults may be injected into different workloads since a single test may not cover all fault locations. 
\tool then performs the fault causality analysis 
to obtain all additional faults triggered. 
We use $I(f_i, t_j)$ to represent a list of additional faults triggered by injecting $f_i$ to $t_j$.

Injections that do not affect system execution 
will be deprioritized automatically.
These non-impactful injections will be clustered together, and assigned the lowest allocation weight in the remaining phases. 
This complements our conservative static filtering of the injection points (\Cref{sec:fia:faults-to-inject}).

\tool uses an IDF-based
(inverse document frequency \cite{schutze2008introduction}, more details in \Cref{app:3pa:idf}) 
clustering algorithm 
to detect \textit{causally equivalent} faults.  
IDF is commonly used in text mining tasks 
for applying weights to the words~\cite{beel2016paper}. 
It excels at reducing noises induced by common words such as ``a'' and ``the''. 
\tool uses IDF to assign a weight to each fault $f$ in the fault corpus $\mathbb F$. 
The intuition is that, similar to common words in text mining, 
if a fault is frequently triggered by various other faults (e.g., inside a utility function), it is less useful for determining the similarity of interferences $I(f_i, t_j)$ from different injections. 
\tool uses such weights to determine the similarity of the interferences caused by different injections. 

Specifically, 
\tool vectorizes each 
$I(f_i, t_{i_1})$ as 
$V(f_i, t_{i_1})=\mathrm{IDF_v}(I(f_i, t_{i_1}), \mathbb{F})$, 
resulting in a real vector of length $n$, with each element ranging from 0 to 1 (i.e., $V \in [0,1]^n$).
\tool then performs 
a hierarchical clustering~\cite{kaufman2009finding} of the faults 
in $\mathbb F$ using the cosine distance~\cite{schutze2008introduction} between all vectorized interferences $V(f_i, t_{i_1})$. 
We select hierarchical clustering due to their explainable nature, 
as required in biology~\cite{krause2005large} and medical~\cite{zhang2017hierarchical} researches. 
We use cosine distance instead of euclidean distance due to their insensitive to vector length (i.e., the number of additional faults triggered in this case). The cosine distance here ranges from 0 to 1 because all IDF vectors are positive.

By the end of phase one, each fault $f_i$ is clustered in to a group $G_j$ with other faults having similar interferences on the system once injected.
This phase 
makes up 25\% of the test budget, 
providing an initial understanding of each fault's interference.

\textbf{Phase Two: Causality Exploration.} 
In phase two, 
\tool distributes test budgets among 
fault clusters $G_i$ in a round-robin manner 
to perform injection and explore their inteferences.  
This approach allocates equal test budgets to fault clusters, 
not individual faults, 
avoiding redundant injections that yield similar interferences. 

In both phase two and three, 
each time a cluster receives a test quota, 
\tool \textbf{randomly} selects one fault from that cluster 
and injects it into a new workload. 
This is to compensate for potential inaccuracies in phase one 
due to limited test budget. 
For example, 
injecting $f_1$ and $f_2$ both causes $f_3$ in phase one, 
but $f_1$ can cause $f_4$ in 
another workload 
which is unfortunately not used in phase one. 
Using random selection, every fault within a cluster have a chance be injected into a new workload.

After completing all the injection runs, \tool performs fault causality analysis again to identify additional faults $I(f_i, t_j)$. 
A second IDF vectorizer is trained with the data from both phases to convert $I(f_i, t_j)$ to $V(f_i, t_j)$, 
which is used in prioritization in the next phase. 

\tool prioritizes to 
inject faults that can cause conditional interferences 
to more workloads, 
in order to extend conditional causal relationships. 
It measures the 
\textit{diversity} 
of the causal consequences (interferences) 
of faults in a cluster 
using 
an intra-cluster 
interference similarity score ($\mathrm{SimScore}(G_i)$). 
\tool prioritizes to 
inject faults from clusters with a lower score, 
which increases the chance for the injection to trigger conditional interferences. 
The score is the average pairwise cosine distance between all vectorized interference results $V(f_i, t_j)$ obtained from the injection experiments conducted in cluster $G_i$ during phase one and two. 
$\mathrm{SimScore}(G_i)$ ranges from 0 to 1. 
A value of 1 indicates that all faults in cluster $G_i$ trigger the same set of additional faults among all injection runs.
The score is then used in phase three.

\textbf{Phase Three: Conditional-Causality-Guided Causality Extension.}
Using the $\mathrm{SimScore}(G_i)$ calculated in phase two, 
\tool performs a weighted random allocation favoring clusters 
with more conditional causal consequences. 
Lower similarity indicates 
more conditional causal consequences 
and thus a higher allocation weight.
The allocation weight for cluster $G_i$ is defined as 
$\max{(\epsilon, 1-\mathrm{SimScore}(G_i))}$.
Each cluster has a minimum weight $\epsilon$ of 0.01, ensuring every cluster receives some budget, even with perfectly matched intra-group interference results.

\textbf{Test Budget Transfer.} 
In phase two and three, 
test quotas are \textit{transferable} between clusters. 
In phase two, 
if a cluster exhausts its (fault, test) combinations before using up its quota, the remaining quota is randomly transferred to a larger cluster for more thorough exploration.
In phase three, any unused budget will be transferred to clusters with a smaller allocation weight.

\textbf{Text Budget Selection.}
We recommend running a minimum of $4 \times |\mathbb{F}|$ tests under the 3PA protocol.
In phase one, each fault needs one test to explore the potential interferences of the injection on the system.
In phase two, each fault is on average injected into two additional test cases, which enables the intra-group similarity calculation. We set a number of 2 instead of 1 in this phase just in case one fault is injected into a similar test workload as phase one. 
In phase three, we naturally allocate each fault one additional test. 

\section{Detect \VCs}
\label{sec:vc-detection}

This sections details how \tool  
stitches causal relationships 
to form fault propagation chains (\Cref{sec:chain-forming}), 
how to avoid incompatible stitching (\Cref{sec:chain-forming:approx-check}), 
and how \tool searches for cycles to detect \vcs (\Cref{sec:guided-search}). 

\subsection{Stitching Causal Relationships}
\label{sec:chain-forming}

\tool stitches causal relationships 
when the resulting fault in one causal relationship 
($f_2$ in $f_1 \rightarrow f_2$ ) 
matches the starting fault in another ($f_2$ in $f_2 \rightarrow f_3$). 
In particular, 
based on the type of the fault ($f_2$) used in stitching, 
\tool can form four types of connections: 
\begin{enumerate}[leftmargin=3ex, noitemsep, topsep=0pt]
\item $f_1 \longEArrow{D/I} f_2 \longAnyInterferenceArrow{I} f_3$: 
The matching fault $f_2$ is an exception or negation. Injecting $f_1$ has an execution trace interference $f_2$ on the system. $f_1$ and $f_3$ can be of any type.

\item $f_1 \longSPlusArrow{D/I} f_2 \longAnyInterferenceArrow{D} f_3$:
The matching fault $f_2$ is a delay on a loop. Injecting $f_1$ has an iteration count interference $f_2$ on the system. $f_1$ and $f_3$ can be of any type. 
For delays in nested loops performing batch processing, there are two variants: 

\begin{enumerate}[leftmargin=3ex]
\item $f_1 \longSPlusArrow{D/I}$ \CFGBackEdge{$f_2$}{$f_2'$}$\longAnyInterferenceArrow{D} f_3$:
The delay $f_2$ propagates to its parent loop ($f_2'$), which is the injected fault of the next injection and causes $f_3$.

\item $f_1 \longSPlusArrow{D/I}$ \CFGBackEdge{$f_2$}{} \CFGEdge{$f_2'$}{$f_2''$}$\longAnyInterferenceArrow{D} f_3$:
The delay $f_2$ propagates to its sibling loop ($f_2''$), which is the injected fault of the next injection and causes $f_3$.
\end{enumerate}
\end{enumerate}

\subsection{Local Compatibility Check}
\label{sec:chain-forming:approx-check}

\begin{figure}
\centering
\footnotesize
\begin{minted}[
  xleftmargin=2ex, xrightmargin=1ex, frame=lines, fontsize=\footnotesize, linenos, 
  autogobble, numbersep=1ex, framesep=1mm, breaklines]
{java}
public ReplicaHandler createTmp(...) {
  do {
    if (current == lastFoundReplica) { ... };
    // ...
    // Fault F2 (Injected)
    if ((current.genStamp() >= b.genStamp()) || !isTransfer)
      // Fault F2 (Triggered by F1 injection)
      throw new ReplicaAlreadyExistsException(...);
  } while (true);
}

class BlockReceiver {
  BlockReceiver(...){
    data.createTmp(...);
  }
}
\end{minted}
\caption{Pseudo-code demonstrating state compatibility. The fault is located inside a loop of \texttt{createTmp()}, which is invoked from \texttt{BlockReceiver()}. Code simplified from \texttt{FsDatasetImpl} in HDFS.}
\label{fig:state-compatibility}
\end{figure}

\noindent
When connecting causal relationships of faults obtained from different tests, \tool performs a compatibility check 
between the tests involved. 
Theoretically, 
to check for 
incompatible propagation activation conditions across tests (and skip stitching), 
a symbolic constraint collection 
(i.e., path condition collection used in symbolic execution techniques~\cite{cadar2008klee,chipounov2012s2e})
should be applied during the fault propagation: 
collect the path conditions $c_1$ and $c_2$
during the propagation of  $f_1 \rightarrow f_2$ in test $t_1$
and $f_2 \rightarrow f_3$ in test $t_2$ respectively, 
and check the satisfiability of their conjunction 
$c_1 \wedge c_2$.

To minimize the overhead, 
\tool approximates this analysis 
by checking whether 1) the \textbf{local} execution trace associated with the fault 
used in stitching ($f_2$) and 2) the call stack 
match with their counterparts in the other test.
In practice, 
we find \tool's strategies successfully eliminates 
most incompatible workloads. 

To facilitate the compatibility check, 
in 
injection runs, 
\tool records call stack and execution trace as follows: 
\begin{enumerate}[wide, labelwidth=!, labelindent=\parindent, noitemsep, topsep=0pt]

\item
\textit{Execution trace recording}. 
\tool records the branch statements and the evaluation results of the conditions 
\textbf{locally} in the fault's enclosing loop or function 
as the path conditions. 
For example, in \Cref{fig:injection-point-exp}, the three \texttt{if}s (line 3, 7, and 15) serve as monitor points for such branch statements. 

\item
\textit{Call stack recording}. 
An error occurred in the same function but at different call sites 
may represent errors for different types of requests, 
which is invalid when stitched together. 
To address this, \tool records the closest two levels of call stack for each loop iteration, exception, and boolean return value (excluding the enclosing function itself). 
This method resembles the  2-call-site sensitivity in pointer analysis, a common method balancing speed and accuracy~\cite{jeon2022return,li2020principled, tan2017efficient}. 
Note that 
the scope of 
execution trace 
and call stack 
used in the comparison 
can both be adjusted. 

\end{enumerate}

\Cref{fig:state-compatibility} shows 
an example for the compatibility check. 
The fault $f_2$ (exception on line 8) used in stitching 
is inside a loop of \texttt{createTmp()}.  
\tool checks the following two things before stitching $f_1 \rightarrow f_2$ and $f_2 \rightarrow f_3$ together: 
1) (call stack) \texttt{create\-Tmp()} must be both invoked from \texttt{Block\-Re\-ceiver()}; 
2) (execution trace) 
the traces of line 3-5 
in the 
fault-happening 
iteration 
of the fault-enclosing loop match between tests. 
If the fault is not within a loop, 
we use the trace of the enclosing function (e.g., \texttt{Block\-Re\-ceiver()}) instead. 
Because delay injection is injected at the beginning of all iterations of a loop, 
\tool conservatively checks 
for matching traces in any loop iterations between tests.

\subsection{Searching for \VCs}
\label{sec:guided-search}

\begin{algorithm}[tb]
\footnotesize
\caption{Parallel Beam Search}\label{alg:beam-algo}
\SetKwProg{Fn}{Function}{}{end}
\KwData{All causal relationships between faults $\mathbb E$}
\KwData{Beam size B}
\KwResult{All \vcs $\mathbb C$}
\BlankLine
\Fn{beamSearch($\mathbb E$, B)}{
    queue $\gets e \in \mathbb E$\;
    $\mathbb C \gets \emptyset$\;
    \While{queue $\neq \emptyset $}{
        next $\gets \emptyset$\;
        \ForPar{$c \in \mathrm{queue}$}
        {
            \ForPar{$e \in \mathbb E$}
            {
                \If{match(c.lastEdge, e)}
                {
                    new $\gets$ c.append(e)\;
                    \eIf{isCycle(new)}
                    {
                        $\mathbb C$.add(new)   
                    }
                    {
                        next.add(new)
                    }
                }
            }
        }
        sort(next)\;
        queue $\gets$ next.subList(0, B)\;
    }
    \Return{$\mathbb C$}\;
}
\Fn{match(edge1, edge2)}{
\Return{edge1.interference == edge2.injectedFault \& isCompatible(edge1.state, edge2.state)}\;
}
\Fn{isCycle(c)}{
\Return{match(c.lastEdge, c.firstEdge)}\;
}
\end{algorithm}

Due to the large search space 
and the need to 
incorporate the local compatibility check, 
\tool uses a customized \textbf{parallel beam search} 
to find cycles, 
as shown in \Cref{alg:beam-algo}. 
The algorithm starts from all causal relationships obtained from the three-phased fault injection, forming propagation chains of length 1 (line 2).
Each while-loop iteration (line 4) represents one-level on the search tree, where one edge \texttt{e} is connected to each propagation chain \texttt{c} in the \texttt{queue} (lines 6-13). Before connection, \tool performs the compatibility check (line 17) between the last edge in the chain and the new edge (line 8). If each chain cycles back to its beginning (line 10), \tool reports a potential \vc.

At each level of the beam search, only $B$ active chains are kept, sorted by the average intra-cluster interference similarity score (\Cref{sec:alloc:3pa}) of the injected faults in the chain (lines 14-15). The intuition is that \tool favors \vcs involving complex error handling logic, which developers might overlook during testing. Suppose each chain $C_i$ has $r$ fault injections $f_{k_1}, f_{k_2}, ..., f_{k_r}$, and each fault $f_{k_j}$ belongs to fault cluster $G_j$. The score used in ranking is defined as: 
$ \mathrm{Score}(C_i) = \Sigma_{j=0}^r \mathrm{SimScore}(G_j) / \lVert C_i \rVert$. 
Chains with lower intra-cluster similarity are kept, as they likely involve conditional causal relationships.

\tool perform the beam search on the entire fault causal space explored by the fault injection. 
Although there is no limit on the number of faults in each chain, chains cannot grow indefinitely as the fault injection runs in the chain must remain compatible.

\textbf{Clustering Reported Cycles.} 
Because in 3PA's phase two and three, 
\tool randomly picks a fault in a cluster of \textit{causally equivalent} faults 
to perform injection, 
our detection algorithm can identify 
equivalent \vcs 
which contains \textit{causally equivalent} faults. 
For example, suppose \tool reports two cycles ``$f_1 \rightarrow f_2 \rightarrow f_1$'' and ``$f_3 \rightarrow f_2 \rightarrow f_3$'', while $f_1$ and $f_3$ are from the same fault cluster $G$, the two cycles reported are likely the same bug due to $f_1$ and $f_3$'s similar interference on the system.
\tool automatically clusters cycles found in the beam search based on the fault clusters (\Cref{sec:alloc:3pa}) involved in the cycle. 

\section{Implementation Details}
\label{sec:impl-detail}

\tool is implemented with over 16,000 lines of Java code, 4,000 lines of Python, and 700 lines of shell scripts. The static analyzer uses WALA~\cite{wala}. The instrumentor and runtime agent are based on a customized version of Byteman~\cite{byteman} for dynamic instrumentation. The test runner uses a modified version of JUnit~\cite{junit}, which works with \tool's instrumentor and the runtime agent to dynamically inject faults before each test. Checkpoint/Restore In Userspace (CRIU)~\cite{criu} snapshots the JVM, speeding up the test initialization.

\para{System-Specific Error Filtering.} 
Except for filtering out 
system-agnostic functions with boolean return value such as Java collection operations 
(\Cref{sec:fia:faults-to-inject}), 
\tool filters out more system-specific errors 
according to the following criteria: 

\begin{enumerate}[wide, labelwidth=!, labelindent=\parindent, noitemsep, topsep=0pt]

\item The boolean return value of the function only involves variables declared as \texttt{final}. 
Such variables are often related to system configurations. 
Configuration errors are not within the scope of this study. 
Instead of hunting for configuration errors, 
\tool searches for fault propagation 
under valid configurations in different workloads. 

\item The return value is constant or never used in the program. 
A negation placed on those functions will not have an impact on the system.

\item The return value is calculated only from primitive-type variables. 
This happens when the function is a utility function 
used in 
implementations of classic algorithms (e.g., \texttt{isSorted()} in a sorting algorithm). 
A negation will cause incorrect calculations, which is not a fault under \tool's scope.
\end{enumerate}

Additional implementation details, 
including a dynamic call graph construction 
and a customized JUnit framework, 
can be found in \Cref{app:impl}.

\section{Evaluation}
\label{sec:eval}

\begin{table}[tb]
\footnotesize
\centering

\caption{Number of injection points, monitor points, and integration tests in each system.}

\begin{tabular}{l|ccccc}
\toprule
\textbf{System} & \textbf{Loop} & \textbf{Exception} & \textbf{Negation} & \textbf{Branch} & \textbf{Test} \\
\midrule
HDFS 2 & 2067 & 2316 & \phantom{0}770 & 26941 & 2674 \\
HDFS 3 & 2736 & 2707 & \phantom{0}974 & 36745 & 4426 \\
HBase & 3227 & 2369 &            1002 & 38192 & 4436 \\
Flink & 2860 & 2619 & \phantom{0}447 & 27947 & 2036 \\
OZone & 1361 & 1395 & \phantom{0}395 & 18212 & 1219 \\
\bottomrule
\end{tabular}
\label{tab:static-result}
\end{table}

We evaluate \tool on the latest versions of \NumOfSysTested popular distributed systems, namely, HDFS 3.4.1~\cite{apache-hdfs3}, HDFS 2.10.2~\cite{apache-hdfs}, HBase 2.6.0~\cite{apache-hbase}, Flink 1.20.0~\cite{apache-flink}, OZone 1.4.0~\cite{apache-ozone}.
The fault injection tests and beam search are performed on two Ubuntu 22.04 servers: one with two Intel Xeon Gold 5220R CPUs and 512 GB of memory, and the other with two AMD EPYC 7313 CPUs and 256 GB of memory. Each profile and injection run is executed in a Docker container with Java 1.8. Each JVM instance is limited to 6 CPU cores and a heap size of 32GB.

\Cref{tab:static-result} summarizes the number of injection points, monitor points, and test workloads identified in each system. 
In all the evaluations, we use a beam size of 5 million chains.

\subsection{New \VCs}
\label{sec:eval:new-bug}

\begin{table*}[tb]
\footnotesize
\centering

\caption{All \NumOfNewVC new \vcs detected by \tool. In general, ``DN'' stands for DataNode, ``NN'' stands for NameNode, and ``IBR'' stands for incremental block report. ``Cycle'' column shows the types of faults in each cycle. ``Alloc.'' column shows the phase number in 3PA protocol that each bug is detected in. A ``\Yes'' in column ``Rnd.?'' means that the bug is detectable under random allocation of test budgets. ``Alt.?'' column shows whether the bug can be triggered by the strategy in \Cref{sec:eval:alt}. ``JIRA\#'' column lists the bug ID in Apache issue tracker.}
\begin{tabular}{llllccccl}
\toprule

\textbf{System} &
\textbf{\#} &
\textbf{Delayed Task} & 
\textbf{Other Faults} &
\textbf{Cycle*} & 
\textbf{Alloc.} & 
\textbf{Rnd.?} & 
\textbf{Alt.?} &
\textbf{JIRA\#}\\
\midrule 

HDFS 2 & 
1) & 
Lease recovery &
IOE in IBR; IOE in write pipeline & 
1D | 2E | 0N & 
1 &
 &
\Yes &
17661\\

 & 
2) & 
Edit log flushing &
IOE in IBR after NN failover & 
1D | 1E | 0N & 
1 &
\Yes &
 &
17836 \\

 & 
3) & 
Block recovery &
IOE in block recovery & 
1D | 1E | 0N & 
1 &
\Yes &
\Yes &
17662\\

 & 
4) & 
Write pipeline &
IOE in block recovery, IBR, and write pipeline & 
1D | 3E | 0N & 
2 &
 &
 &
17837 \\

 & 
5) & 
Block cache &
IOE in write pipeline; DN timeout & 
1D | 1E | 1N & 
2 &
 &
 &
17660 \\ 

 & 
6) & 
IBR &
IOE in IBR & 
1D | 1E | 0N & 
3 &
 &
 &
17780 \\
\midrule 

HDFS 3 & 
1) & 
Block deletion &
IOE in write pipeline; DN timeout & 
1D | 1E | 1N & 
2 &
\Yes &
 &
17838 \\ 
 
 & 
2) & 
Block reconstruction; IBR &
DN timeout; IOE in replication & 
2D | 1E | 1N & 
3 &
 &
 &
17782 \\ 
\midrule 

HBase & 
1) & 
Write ahead log (WAL) &
Premature \texttt{EndOfFile} in WAL & 
1D | 0E | 1N & 
1 &
\Yes &
\Yes &
29600\\

 & 
2) & 
Region assignment &
IOE in assignment RPC; Node exclusion & 
1D | 1E | 1N & 
3 &
 &
 &
29006 \\ 
\midrule 

Flink & 
1) & 
Task worker &
Head task failure; Sink task cancellation & 
1D | 2E | 0N & 
1 &
\Yes &
 &
38367 \\

 & 
2) & 
Aggregation task &
Task state transition failure; Barrier task failure & 
1D | 2E | 0N & 
2 &
\Yes &
 &
38368 \\
\midrule

OZone & 
1) & 
Container report queue &
Event queue dispatch failure & 
1D | 0E | 1N & 
1 &
 &
 &
13020\\

 & 
2) & 
Heartbeat handling &
IOE in pipeline construction; Pipeline unhealthy & 
1D | 1E | 1N & 
2 &
\Yes &
\Yes &
$11856_1$\\

 & 
3) & 
Replication command handling &
IOE in replication and pipeline construction & 
1D | 2E | 0N & 
3 &
 &
 &
$11856_2$ \\
 
\bottomrule

\end{tabular}

\hspace{48em}\scriptsize{*D: Delay; E: Exception; N: Negation}

\label{tab:bug-detected}
\end{table*}

\Cref{tab:bug-detected} shows the \NumOfNewVC new \vcs detected by \tool in all \NumOfSysTested distributed systems, many of which involve retries and the failure recovery logic. This aligns with the findings of \citet{qian2023vicious} on \vcs.  
\NUMOfConfirmedVC of them have been confirmed, with \NumOfFixedVC fixed. 
Two bugs detected in HDFS 2 are also detected in HDFS 3, we do not list them in \Cref{tab:bug-detected} due to duplication. 

\Cref{tab:bug-detected} also details the characteristics of the fault injection runs for each bug. Column ``Cycle'' shows the number of faults injected of each type.
Column ``Alloc.'' shows the test budget allocation phase after which all the causal relationships of each bug are discovered. All three phases of the 3PA protocol contributes to detecting new bugs. 

To further demonstrate 3PA protocol's effectiveness, we compare it with a random allocation protocol, running the same number of fault injection runs as the 3PA protocol with randomly selected (fault, test) combinations. A ``\Yes'' in column ``Rnd.?'' of \Cref{tab:bug-detected}  means that the bug can be detected using random allocation. The random selection allocation protocol, albeit effective in some systems (i.e., Flink), generally under-performs the 3PA protocol in \tool. 

Most new failures detected by \tool require only one delay injection, with varying numbers of exception and negation injections. 
This is in line with the bug dataset provided by \citet{qian2023vicious}, where majority of the cases only requires a single contention to be triggered. 

\Vcs reported by \tool that involve multiple delay injections generally contribute to the false positive cases. 
However, most of the false positive cases 
are valid fault propagations, 
but categorized as false positive 
because they are known (and accepted by the developer) 
contentions between operations in the system 
(e.g., contention between the HDFS clients performing read and write). 
Details will be discussed in the \Cref{sec:eval:fp}.

\subsection{Comparison with Alternative Strategy}
\label{sec:eval:alt}

We compare \tool with a naive strategy that 
injects a single fault into a system and monitors whether it causes itself 
(e.g., delays a single loop in the system and monitors if its iterations increase). 
This is to evaluate whether the triggering conditions of each detected bug span multiple tests.

Column ``Alt.?'' in \Cref{tab:bug-detected} shows that about \NoNaivePctg (\NumOfNewVCNaive out of \NumOfNewVC) of the \vcs detected by \tool cannot be triggered by the naive strategy. 
This underscores the effectiveness of \tool's causal stitching, the complexity needed to trigger these failures, and the limitations of single-fault injection tools. Even when the naive strategy works, \tool offers detailed insights into the cycle's behavior.

\subsubsection{Comparison with Existing Fuzzing Techniques}

In addition to the naive strategy above, we compare \tool with blackbox fuzzing techniques. Specifically, we compare with Jepsen~\cite{jepsenio} and its Python reimplementation, Blockade~\cite{blockadetest}. Jepsen is a Clojure-based blackbox distributed system fuzzer, widely being used in the security research community~\cite{meng2023greybox,zou2025blackbox} for distributed system testing.

We apply Jepsen on Flink~\cite{jepsenflink} and Blockade on OZone~\cite{blockaeozone} with their existing test cases. Due to the lack of Jepsen or Blockade integration and test workload, HBase and both versions of HDFS are not included in this comparison.

The result shows that none of the 
\vcs detected by \tool can be detected by fuzzing-based techniques. This highlights the necessity of causal stitching and fine-grained fault injection in \tool for detecting \vcs. 

\subsection{Case Study}
\label{sec:eval:case-study}

In this section, we present case studies on two of the confirmed \vcs.

\subsubsection{Region Deployment Retry in HBase}

\begin{figure}
    \centering
    \includegraphics[width=0.8\columnwidth]{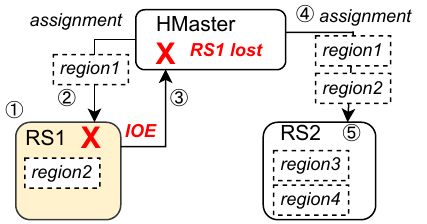}
    \caption{The case study. ``RS'' stands for RegionServer, and ``IOE'' stands for IOException. The yellow box indicates a delay.}
    \label{fig:motivating-exp}
\end{figure}

The second HBase bug in \Cref{tab:bug-detected} is a \vc 
similar to the AWS motivating example in \Cref{sec:intro}. 
HBase is a distributed database system 
that divides tables into regions managed by multiple RegionServers (RS) under master nodes.

At high level, RSes in a heavily loaded cluster may throw IOExceptions (IOEs) when handling RPC requests. The \vc happens in a cluster with a lot of write and table creation requests, which place significant load on the RSes managing these tables (\circledtext{1} in \Cref{fig:motivating-exp}). Some region assignment operations on these RSes may time out and throw IOEs (\circledtext{2}). 
When an RS throws an IOE, it is excluded from the master node's \texttt{Favored\-Stochastic\-Ba\-lan\-cer}, which requires at least three live RSes to function properly. A reduced number of live RSes causes the load balancer to fail (\circledtext{3}). 
An improper handling logic blindly retries the assignment indefinitely (\circledtext{4}), further increasing the load on the RSes (\circledtext{5}). This creates a self-sustaining cycle, leaving the cluster unable to process additional region assignment requests.

To trigger this bug in a testing environment with a single-fault (i.e., delay) injection, we need a workload with the following conditions:
\begin{enumerate}[leftmargin=5ex, noitemsep, topsep=0pt]
\item It contains many region assignment requests. 

\item A cluster configuration prone to be overloaded (3 nodes). 

\item Load balancer configured to \texttt{F.S.Balancer}. 

\item A workload long enough to observe the cycle. 
\end{enumerate}

However, no single test in HDFS satisfies 
all these conditions (as shown in \Cref{sec:eval:alt}), 
and 
\tool detects this cycle by injecting one delay, one exception, and one return-value negation into 
\textbf{three separate tests} $t_1$, $t_2$, and $t_3$. 

First, the delay is injected into the region deployment loop in 
$t_1$
with multiple table creation and clones (cf. \circledtext{1} in \Cref{fig:motivating-exp}). This overloads the cluster, causing IOEs in the region assignment request handler (cf. \circledtext{2}). 
This causal relationship is only detected in 
$t_1$, 
because other tests only exercise limited number of region assignments. 

In the second experiment, 
the same IOE is injected into the region assignment request handler 
in a test ($t_2$) for RS fault tolerance, 
triggering a negated return value in the load balancer's status checker (cf. \circledtext{3}). 
This causal relationship is only detected 
in $t_2$, 
because $t_2$ is the only test that 
uses 
\texttt{Favored\-Stochastic\-Balancer} 
with a cluster of 3 nodes, 
and 
the error detector \texttt{canPlaceFavoredNodes()} 
in \texttt{Favored\-Stochastic\-Balancer} 
only runs into an error (negation) 
if the cluster has fewer than 3 nodes 
(e.g., $t_3$ uses \texttt{Favored\-Stochastic\-Balancer} but has 5 nodes).

Finally, in the third experiment ($t_3$), the return-value negation is injected into the load balancer during a test for the balancer itself, increasing the iterations of the region deployment loop (cf. \circledtext{4} and \circledtext{5}), indicating an increase in the workload being handled that can potentially cause delays. 
This causal relationship is only observed in 
$t_3$, 
because 
it is the only test that uses 
\texttt{Favored\-Stochastic\-Balancer} configuration 
and has a long enough workload to observe the increased iterations 
(not in $t_2$ because it exits prematurely after the IOE). 

\subsubsection{Bypassed IBR Throttling in HDFS}

The sixth bug detected on HDFS 2 is a logic error that a failed 
incremental block report (IBR) is retried immediately at the next heartbeat, 
incorrectly ignoring the configured IBR interval. 
The unconstrained, immediate retry of IBR 
can cause a \vc if the original IBR failure is 
due to a server overload. 
This happens when the cluster 
is under high user traffic with many modifications, 
which triggers many large-sized IBRs.

\tool detects this bug by linking two injections in two different tests $t_1$ and $t_2$: 1) injecting an IBR processing delay into a workload with over 5,000 blocks to test the load balancer, and 2) injecting an RPC exception for IBR into a workload testing the IBR report interval configuration. 

In fault injection using $t_1$, the RPC exception is observed, but not an increase in IBR count. 
Because $t_2$ is the only one with IBR throttling configured. 
Without it, IBRs are sent with every heartbeat. 
As a result, 
although IBR failures occur in the first workload, they are still sent at the original frequency and cannot be detected by execution trace comparison. 

In fault injection using $t_2$, 
the IBR increment is observed 
after the RPC exception, indicating a potential delay. 
However, $t_2$ involves only 8 file blocks 
compared to 5,000 in $t_1$, 
making IBR processing delays less likely to cause timeouts.

\subsection{Cycle Clustering and False Positive Rate}
\label{sec:eval:fp}

\begin{table}[tbp]
\centering
\footnotesize
\setlength{\tabcolsep}{3pt}

\caption{Number of \vcs reported by \tool and the clustering results. Column ``System'' is the targeted distributed system. Columns ``Cycle'', ``Cluster'', and ``TP'' shows the number of cycles reported, distinct cycle clusters, and true positive clusters respectively. The numbers outside the parentheses are from a beam search with unlimited number of injected faults, while the numbers inside are from a beam search limiting to one delay injection.}
\begin{tabular}{lccc||lccc}
\toprule
\textbf{System} & \textbf{Cycle} & \textbf{Cluster} & \textbf{TP} &
\textbf{System} & \textbf{Cycle} & \textbf{Cluster} & \textbf{TP}\\
\midrule 

HDFS 2 & \phantom{0}38 (23) & 15 \phantom{0}(9) & 6 (6) &

Flink & 48 (27) & 35 (17) & \phantom{0}2 (2)\phantom{x} \\

HDFS 3* & 149 (59) & 36 (14) & 4 (3) &

OZone & 29 (17) & 11 \phantom{0}(7) & \phantom{0}3 (3)\phantom{x} \\

HBase & \phantom{0}72 (26) & 24 \phantom{0}(7) & 2 (2) &

\textbf{Total}* & & \multicolumn{2}{r}{\textbf{\NumOfNewVCAllowDupe (\NumOfNewVCOneDelayAllowDupe)}} \\

\bottomrule
\end{tabular}

\hspace{9em}\scriptsize{*Including two duplicated clusters also detected in HDFS 2}

\label{tab:cluster-result}
\end{table}

\Cref{tab:cluster-result} shows the number of \vcs, distinct failure clusters, and the true positive reported by \tool on each system. 
The numbers outside the parentheses are from a beam search without an upper bound on the number of injected faults, while the numbers inside the parentheses are from a beam search for cycles with at most one delay injection but unlimited exceptions or negations.

Limiting the beam search to one delay injection generally reduces false positives while still identifying most failures. Users can adjust the number of injected faults to balance between accuracy and completeness.

\subsubsection{Effectiveness of Cycle Clustering}

We manually inspect all the \vcs and the \textit{causally equivalent fault} clusters on HDFS 2 and confirm that the clustering algorithm effectively groups similar cycles together. Examples of the clustered cycles involve faults from 1) the write pipeline, 2) DataNode command generation and processing. 

However, we do observe false positives in the clustering results due to 1) non-determinism and 2) insufficient test execution. For example, a fault cluster in a HDFS run mixed four correctly-clustered data-race faults with two false positives: a failed security check whose unique causal consequence would have been revealed with more test cases, and a logging error clustered due to non-determinism. This could cause missed \vcs in detection if the false positive fault is selected during the 3PA allocation, though their impact is limited as long as the majority are clustered correctly.

The higher numbers of cycles and clusters reported in HDFS 3 are due to the extensive usage of asynchronous tasks and event queues, 
where 
errors in the issuer 
and the executor of asynchronous tasks 
are processed by different error handlers. 
Compared to synchronous execution, 
this 
increases the number of error handlers, 
resulting in 
a larger number of 
cycles and fault clusters.

\subsubsection{False Positive Analysis}

The false positive clusters reported by \tool are mainly due to the following reasons:
\begin{enumerate}[wide, labelwidth=!, labelindent=\parindent, noitemsep, topsep=0pt]
\item 
In cycle searches with unlimited fault injections, many false positives (about 70\% on HDFS 2) are due to ``expected'' contentions, such as between HDFS clients with heavy read and write operations. These cases require solutions such as throttling or capacity increases. 
We do not consider them as true positives. 

\item 
Increased loop iterations don't always indicate delays, especially if throttling mechanisms are already in place. These mechanisms are often implemented ad-hoc, making them difficult for static analyzers to identify.

\end{enumerate}

From our experience, 
the false positives 
due to the above reasons can be 
filtered out manually 
relatively easily, 
given adequate understanding of the system under test. 

Theoretically, \tool could still report a \vc involving fault propagation chains from tests with contradictory conditions after our local compatibility check. However, we do not observe such cases in the clusters reported by \tool.

\subsection{Performance Overhead}
\label{sec:eval:perf}

\tool's instrumentation and monitoring introduces an average of 185\% runtime overhead in the profile run, ranging from 63\% to 376\%. This overhead is primarily caused by branch tracing and call stack recording, 
and can be further reduced using hardware-based tracing such as IntelPT~\cite{henschelintelpt}. 
While sampling can be used to reduce the runtime overhead, we opt not to implement sampling in branch tracing to maximize the detection accuracy as \tool is intended to be deployed in the testing environment instead of the production environment.

\subsection{Effort of Applying \tool to a New System}

\tool is designed to be system-agonistic and requires as little input from users as possible. Applying \tool to a new system only requires two steps: 1) compiling the target system, and 2) setting up the test environment with system dependencies. Two graduate students who have no knowledge of the tool can independently apply \tool to a new system within half a day without consulting the authors. 
\section{Discussion}

In this section, we discuss \tool's generality, its usage in real-world scenarios, and lessons learned from the \vcs detected.

\subsection{Generality of \tool}

\para{Across distributed system components.} 
\tool can identify causal relationships across components of distributed systems 
(currently supports JVM-based components).
For example, \tool can identify causal relationships between faults in HBase and its underlying HDFS, provided that integration tests driving multiple system components are available. 

\para{Generality of faults.}
\tool's fine-grained fault injection (\Cref{sec:fia:faults-to-inject}) is both representative and general.
First, we achieve completeness in exception injection by targeting all \texttt{throw} statements, which are then conservatively pruned via a rule-based approach. Second, we also achieve completeness in contention injection for \vcs by simulating their common cause -- workload-induced contention~\cite{huang2022metastable, qian2023vicious} -- with spinning delay.
Third, we make a practical tradeoff with the system-specific error injection by using a heuristic (i.e., targeting boolean-returning functions) to maintain generality. While this heuristic may miss some error detectors implemented as branch conditions, such cases are often captured by exception injections if they guard \texttt{throw} statements. Conversely, a fully complete approach would require  input from developers to distinguish normal branch conditions from the remaining system-specific error detectors.

\subsection{Using \tool in Real-World Scenarios}

We expect \tool to be applied 
before releases 
to catch potential \vcs due to its exhaustive testing approach. 
To adapt \tool for regular regression testing, \tool's budget allocation algorithm can be refined to focus on faults in the classes or packages that 1) have been involved in prior failures or 2) involve heavy code changes and newly introduced test workloads. However, this remains a future direction that requires thorough evaluation.

\subsection{Lessons Learned}

Several ways can be applied to reduce the possibility of encountering \vcs in real-world distributed systems. First, workload should never be piggybacked on critical requests (e.g., heartbeat or failure detectors). A surge in the user traffic can easily drive the system into a cascading failure. Second, distributed systems should properly implement throttling mechanisms by proactively monitoring current system load such as the queue length. In addition to dropping requests at the server side, clients should be properly informed to refrain themselves from retrying to avoid a retry storm. Third, asynchronous requests can be used with priority queues to ensure that critical operations are handled first in a unusual traffic spike.

\section{Related Work}

\para{Cascading Failure.} 
\tool gains insight from the 
work of \citet{qian2023vicious} and \citet{huang2022metastable} on \vcs in distributed systems. Following their work, diagnosis~\cite{harsh2023murphy,li2024exchain, sruthi2024demystifying}
and mitigation~\cite{li2023towards, meza2023defcon} techniques 
have been proposed and studied. 
However, none of them focuses on 
exposing 
\vcs caused 
using testing or fault injection. 
\citet{li2018pcatch} 
proposes a technique to detect 
\textit{performance cascading failures}, 
a subtype of cascading failures that 
only involves performance interference 
by tracking a variant of happens-before relationship causality. 
In comparison, 
\tool detects fault propagation involving both performance interference and functional interference 
through counterfactual causality analysis. 

Another line of work uses queuing models~\cite{habibi2023msf} and Markov chains~\cite{isaacs2025analyzing, isaacs2025formal} to analyze the behavior of request-response server systems and identify potential metastable failures. Such approaches require developers to provide multiple layers of specifications, including CTMC (continuous time Markov chain) models and DES (discrete event simulator) models. 
In comparison, \tool works on the bytecode of the system directly and is not limited to request-response server systems.

\para{Automatic Bug Detection and Diagnosis in Distributed Systems.} Recently, there has been a rise in the focus on automatically detecting and diagnosing failures in distributed systems, including performance cascading bugs~\cite{li2018pcatch}, exception-dependent failures~\cite{li2024exchain}, upgrade failures~\cite{zhang2021understanding}, timeout bugs \cite{chen2024chronos}, and partial failures~\cite{wu2024efficient}. Compared with their work, \tool's causal stitching links multiple faults injected in to the system. None of the prior work can provide detailed insight into the root causes as well as the propagation chains of the \vcs.

\para{Fault Injection.} 
Traditionally, 
fault injection techniques
\cite{hsueh1997fault,
marinescu2009lfi, 
natella2016assessing,bai2016testing,bai2016mining}
focus on 
single-node system. 
Recently, 
it has become a popular~\cite{alquraan2018analysis, chen2020cofi, ganesan2017redundancy, gunawi2011fate, lu2019crashtuner, 
marinescu2009lfi,
alvaro2015lineage,
joshi2011prefail,
ju2013fault,
mohan2018finding,
li2019efficient,
sun2022automatic,
meiklejohn2021service,
zhang20213milebeach,
chen2023push,
jepsenio} 
technique in testing distributed systems. Compared with \tool, many of them inject coarse-grained external faults such as node failures and network partitions to expose crash recovery bugs. Although more advanced techniques exist, such as injecting faults during variable accesses~\cite{lu2019crashtuner}, none of them are able to detect and expose the entire causal chain of \vcs. \tool performs fine-grained injection of delay, exception, and return value negation into the system. Combined with the fault causality analysis, \tool is the only tool that can detect and diagnose \vcs in distributed systems.

\para{Fuzzing.}
Fuzzing is a software testing technique that feeds programs with large volumes of random
inputs to uncover bugs with predefined oracles such as crashes. Prior work mainly applies fuzzing techniques to single-node systems and programs~\cite{jiang2020fuzzing, jiang2022context, xie2022rozz, gao2025detecting, jiang2023dynsql, bai2024multi,bai2023testing, afl, AFLplusplus-Woot20, bohme2017directed, dinesh2020retrowrite}, such as operating systems~\cite{vyukov2015syzkaller,gong2025snowplow,sun2021healer,jiang2019fuzzing}, network protocols~\cite{zou2021tcp,pham2020aflnet}, and filesystems~\cite{jiang2022context,kim2019finding}. Recently, fuzzing has also been applied to test distributed systems~\cite{crashfuzz, meng2023greybox, zou2025blackbox} and shows promising results.

\tool complements existing black-/grey-box fuzzing techniques. While fuzzing explores new behaviors by mutating test inputs (e.g., message sequences, code coverage), \tool extends fault causality chains, which can be used as feedback information for existing fuzzers. For instance, when \tool stitches $f_1 \rightarrow f_2$ in $t_1$ and $f_2 \rightarrow f_3$ in $t_2$ to $f_1 \rightarrow f_2 \rightarrow f_3$, such a causality chain can guide the fuzzers to mutate $t_1$ and $t_2$ together (e.g., merging workload in both tests), helping triggering \vcs.

\section{Conclusion}

We propose \tool, a fault-injection-based testing and detection framework for \vcs in distributed systems. \tool leverages the novel idea of causal stitching to expose the complex fault propagation chains and detects potential \vcs due to incorrect program logic. We evaluate \tool on \NumOfSysTested popular distributed systems and find \NumOfNewVC new \vcs, \NumOfConfirmedVC of which have been confirmed with \NumOfFixedVC fixed by developers. \tool is effective with little burden on targeted users.

\begin{acks}
The authors thank Jia-Ju Bai, our shepherd, and anonymous reviewers for their constructive comments. This research is partially 
supported by NSF 1901242, 2006688, and 2300562.
Any opinions, findings, and conclusions in this paper are those of the authors 
only and do not necessarily reflect the views of the sponsors.
\end{acks}

\clearpage


\bibliographystyle{ACM-Reference-Format.bst}
\bibliography{paper}

\clearpage

\appendix

\section{Details on 3PA Protocol}
\label{app:3pa}

Suppose there are $n$ possible fault locations in the system, namely $f_1, f_2, ..., f_n$, making up the fault space $\mathbb F$. Injecting $f_i$ into test $t_j$ causes a $m$ additional faults: 
\begin{equation} 
I(f_i, t_j)=[f_{{k_1}}, f_{{k_2}}, ..., f_{{k_m}}]\quad k_x\in[1..n], x \in [1..m]
\end{equation}

\subsection{IDF Vectorization}
\label{app:3pa:idf}

Each $I(f_i, t_{i_1})$ is vectorized using the inverse document frequency (IDF)~\cite{schutze2008introduction}, resulting in a real vector of length $n$, with each element ranging from 0 to 1. Vectorization is a common technique before applying clustering algorithms~\cite{schutze2008introduction}.
\begin{equation}
V(f_i, t_j)=\mathrm{IDF_v}(I(f_i, t_j), \mathbb{F}) \quad \in [0,1]^n
\end{equation}

IDF measures how often a fault is triggered in all the fault injection runs. Formally, it is defined~\cite{schutze2008introduction} as:
\begin{equation}
\mathrm{IDF}(f, \mathbb F) = \log \frac{1 + N}{1 + N_f}
\end{equation}

$N$ is the total number of fault injection experiments, and $N_f$ is the number of experiments that triggers the additional fault $f$. The plus one smooths the IDF values and prevents zero divisions.

To vectorize $I(f_i, t_j)$ using IDF, each triggered fault is replaced with its IDF value, and other elements in $V(f_i, t_j)$ are set to zero. L2 normalization is applied to the final vector. Formally, the vectorization is defined as:
\begin{equation}
\begin{aligned}
V(f_i, t_j) &= \mathrm{IDF_v}(I(f_i, t_j), \mathbb{F}) = \frac{(v_1, v_2, ..., v_n)}{\lVert (v_1, v_2, ..., v_n) \rVert} \\
v_{x} &= 
\begin{cases}
    \mathrm{IDF}(f_x, \mathbb F), & \mathrm{if} \; f_x \in I(f_i, t_j) \\
    0,                            & \mathrm{otherwise}
\end{cases} \\
\mathrm{where} \quad x &\in [1..n]
\end{aligned}
\end{equation}

\subsection{Phase One}

In phase one of the 3PA protocol, \tool injects each fault $f_i$ into workload $t_{i_1}$ that reaches $f_i$'s program location 
and has the highest code coverage. 
We perform a hierarchical clustering~\cite{kaufman2009finding} of the faults in $\mathbb F$ using the cosine distance $\mathrm{D_c}$ between all vectorized interferences $V(F_i, T_{i_1})$, as defined below:
\begin{equation}
\begin{aligned}
\mathrm{D_c}(V(f_i, t_{i_1}), V(f_j, Tt_{j_1})) &= 1 - \frac{V(f_i, t_{i_1}) \cdot V(f_j, t_{j_1})}{\lVert V(f_i, t_{i_1}) \rVert \lVert V(f_j, t_{j_1}) \rVert } \\
\mathrm{where} \quad i, j &\in [1..n], i \neq j
\end{aligned}
\end{equation}

By the end of phase one, each fault $f_i$ is clustered in to a group $G_j$ with other faults having similar interferences on the system once injected.

\subsection{Phase Two}

In phase two of the 3PA protocol, an intra-cluster interference similarity score is calculated for each cluster $G_i$.
This score is the average pairwise cosine distance of all vectorized interference results within the cluster.
That is, suppose $G_i$ has $p$ faults $f_{i_1}, f_{i_2}, ..., f_{i_p}$, $f_{i_k}$ is injected into $q_k$ different workloads $t_1^{i_k}, t_2^{i_k}, ..., t_{q_k}^{i_k}$, cluster $G_i$ has $\Sigma_{k=1}^p q_k$ vectorized interference results $V(f_i, t_j)$, the similarity score is defined as:
\begin{equation}
\begin{aligned}
\mathrm{SimScore}(G_i) &= 1-\overline{\mathrm{D_c}(V(f_a, t_x^a), V(f_b, t_y^b)} \; \in [0,1] \\
\mathrm{where} \quad a, b &\in [1..p], x \in [1..q_a], y \in [1..q_b], a \neq b
\end{aligned}
\end{equation}

$\mathrm{SimScore}(G_i)$ ranges from 0 to 1. A value of 1 indicates that all faults in cluster $G_i$ triggers the same set of additional faults among all injection runs.

\subsection{Phase Three}

In phase three of the 3PA protocol, 
the budget allocation weight for cluster $G_i$ is defined as: 
\begin{equation}
\mathrm{W}(G_i)=\max{(\epsilon, 1-\mathrm{SimScore}(G_i))} \quad \in [0,1]
\end{equation}

Each group has a minimum weight $\epsilon$ of 0.01, ensuring every cluster receives some budget, even with perfectly matched intra-group interference results.

\section{Additional Implementation Details}
\label{app:impl}

\subsection{Dynamic Call Graph Collection}

Loop scalability analysis in \tool (\Cref{sec:fia:faults-to-inject}) requires a call graph to identify the functions reachable from each loop.
During development, we find that WALA's static call graph struggles with polymorphism. More accurate algorithms such as 2-CFA~\cite{shivers1988control} do not scale well for large systems, such as HDFS with over 359,000 lines of code.
To address this, we uses \texttt{async-profiler}~\cite{async-prof}'s CPU sampler alongside \tool's tracing capabilities to collect runtime stack snapshots, from which a dynamic call graph is reconstructed.

\subsection{Parameterized JUnit Test} 

A challenge in test execution is the wide use of parameterized JUnit tests~\cite{junit-parameterized-test}, which reuses the test methods with different input parameters, often generated dynamically.
\tool uses a modified version of JUnit that skips the test body execution and focuses solely on parameter generation. A custom test filter intercepts this process and captures parameters for testing.

\section{Artifact Appendix} 

This artifact contains a minimum working example of the \tool. 

\subsection{Description \& Requirements}

\subsubsection{How to access}

The code can be downloaded at \url{https://doi.org/10.5281/zenodo.17049891}.

The Zenodo artifact contains one file, \texttt{CSnake-AE.tar.zst}, it contains a folder \texttt{CSnake-AE} with a \texttt{README.md} inside. The readme file contains all the necessary steps for the artifact evaluation process.

It is mandatory that the file is downloaded into \texttt{``/''} under Linux and extracted there. In other words, the code will be located at \texttt{/CSnake-AE}.

\subsubsection{Hardware dependencies}

Detailed requirements are listed inside \texttt{README.md}. We recommend using a machine with at least 256GB of memory and 50GB of SSD storage space. 

\subsubsection{Software dependencies}

We list detailed commands for installing all the software dependencies in \texttt{README.md}.

\subsubsection{Benchmarks} 

None.

\subsection{Set-up}

See \texttt{README.md} for details.

\subsection{Evaluation workflow}

\texttt{README.md} contains nine steps of demonstrating the functionality of \tool. 

\tool's static analyzer is executed in Step 1.
\tool's profile run of integration tests is executed in Step 2--4. Step 5 analyzes the output and prepare \tool for fault injection runs. Step 6 and 7 executes the fault injection run. Meanwhile, our runtime agents for fault injection and monitoring is exercised. Step 8 performs the fault causality analysis. Step 9 runs the bug detector, performing the local compatibility check and parallel beam search.

\end{document}